УДК XXXXXXX
# Анализ представления в SIMBAD данных для звёзд каталога GTSh10

*А.А. Шляпников* [1]

[1] ФГБУН «Крымская астрофизическая обсерватория РАН», Научный, Крым, Россия, 98409
*aas@craocrime.ru*

**Аннотация.** Представлен анализ данных о звёздных величинах объектов из каталога GTSh10 в базе данных SIMBAD. Для исключения неоднозначности относительно состояния блеска проведено сравнение звёздных величин из SIMBAD и каталогов NOMAD, SDSS и VSX. Показано, что для некоторых объектов в SIMBAD представлены звёздные величины в состоянии высокой активности (возможно вспышечной). Обнаруженные несоответствия необходимо учитывать при планировании оригинальных наблюдений и сравнении данных GTSh10 с вновь появляющимися обзорами объектов всего неба.

ANALYSIS OF REPRESENTING DATA FOR STARS FROM THE GTSh10 CATALOGUE IN SIMBAD, *by A.A. Shlyapnikov.* We present an analysis of data on stellar magnitudes of objects from the GTSh10 catalogue in the SIMBAD database. To eliminate ambiguity about the state of brightness, a comparison of stellar magnitudes from SIMBAD and NOMAD, SDSS and VSX catalogues is carried out. It is shown that for some objects in SIMBAD stellar magnitudes are presented in a state of high activity (possibly flare). The detected inconsistencies must be taken into account when planning original observations and comparing the GTSh10 data with the newly appearing surveys of the sky.

**Ключевые слова:** каталоги, фотометрия, анализ

## 1 Введение

При подготовке каталога звёзд с активностью солнечного типа - GTSh10 (Гершберг, 2011) возникла неоднозначность при добавлении в него значений звёздных величин вспыхивающих звёзд. Одной из задач составления GTSh10 было обеспечение возможности планирования наблюдений, включенных в каталог объектов. Это предполагало, что в GTSh10 будут представлены звёздные величины вспыхивающих звёзд в неактивной фазе (не во время вспышек).

Однако, поскольку основу GTSh10 в части идентификации, координат и звёздных величин вошедших объектов составила информация, содержащаяся в базе данных SIMBAD (Женова, 2005), было обнаружено несоответствие приведённых в ней звёздных величин вспыхивающих звёзд данным, содержащимся в других каталогах. Учитывая подготовку дополнений к GTSh10, которые составят основу нового каталога звёзд с активностью солнечного типа, проведённое исследование приобретает особую актуальность.

В данной статье представлен анализ данных о звёздных величинах, содержащихся в базе данных SIMBAD и каталогах NOMAD, SDSS и VSX.

## 2 Сравнение данных SIMBAD и NOMAD

Зависимость между звёздными величинами $B$, $V$ и $J$ из SIMBAD и каталога NOMAD (Захариас, 2005) для 1056 объектов представлена на рис. 1. Видно, что для большинства звёзд слабее $14^m$ имеет место явное несоответствие между значениями блеска. Для более детального анализа были отобраны лишь 828 звёзд с $B$ величинами, которые в большем количестве представлены в SIMBAD (рис. 2). Рисунок иллюстрирует, что для большинства звёзд значения их блеска в SIMBAD завышено.

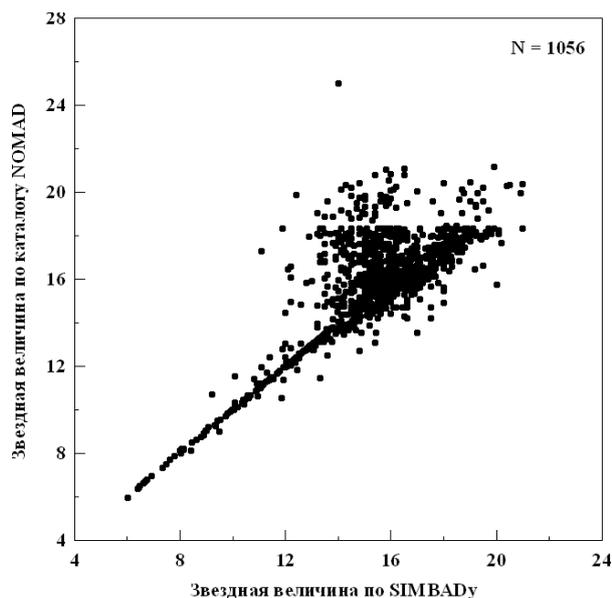

Рис. 1.

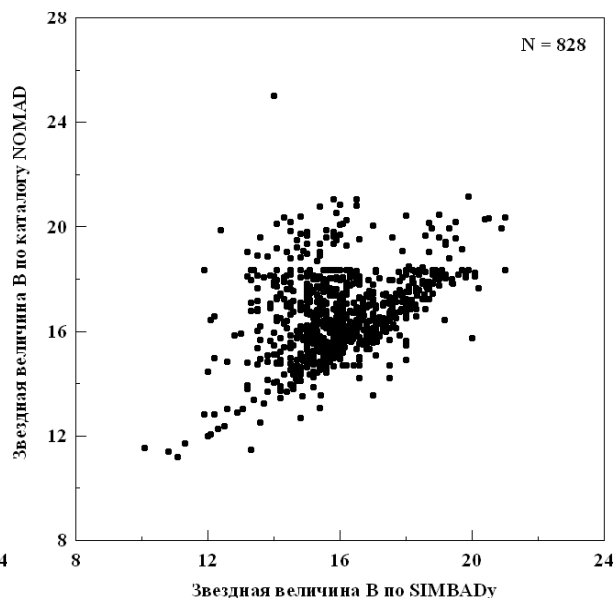

Рис. 2

## 3 Сравнение данных NOMAD и SDSS

Для большей уверенности в данном выводе была произведена выборка вспыхивающих звёзд GTSh10, найденных в NOMAD, из каталога SDSS 7-й реализации (Абазаян, 2008). При сравнении звёздных величин 39 объектов, присутствующих в обоих каталогах подтвердилось предположение об указании в SIMBAD большего блеска для объектов в сравнении с другими каталогами.

Лишь один объект показывает значительное удаление от близкой к линейной зависимости между $B$ величинами, взятыми из каталогов NOMAD и SDSS (рис. 3). И совершенно отсутствует какая-либо зависимость в случае сравнения $B$ величин SIMBAD - NOMAD (рис. 4) и SIMBAD - SDSS (рис. 5).

Коэффициент корреляции данных, представленных на рис. 3, составляет 0.96 (без значительно удаленной точки), в то время как

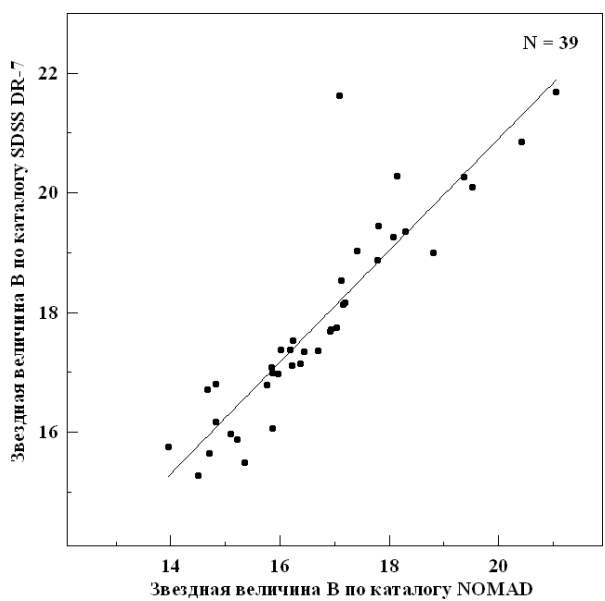

Рис. 3.

для данных, представленных рис. 4 и рис. 5, 0.40 и 0.36 – соответственно. Большая асимметрия в разбросе точек направленная в верхнюю часть рисунка и некоторое подобие нижней огибающей, также указывают на завышенный блеск для вспыхивающих звёзд, приведенный в SIMBAD.

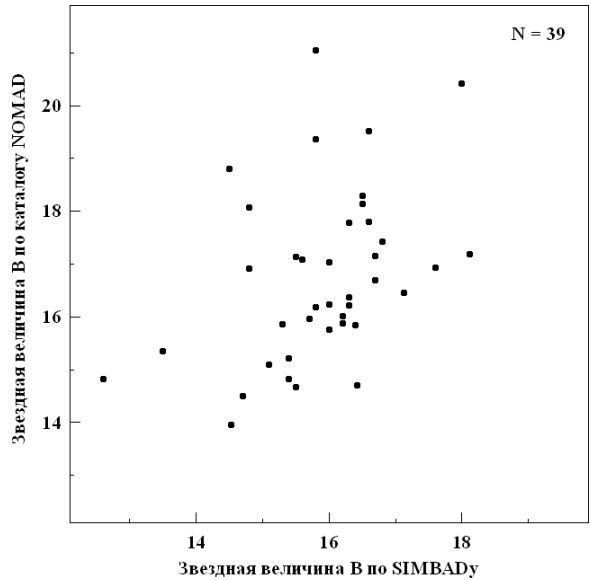

Рис. 4.

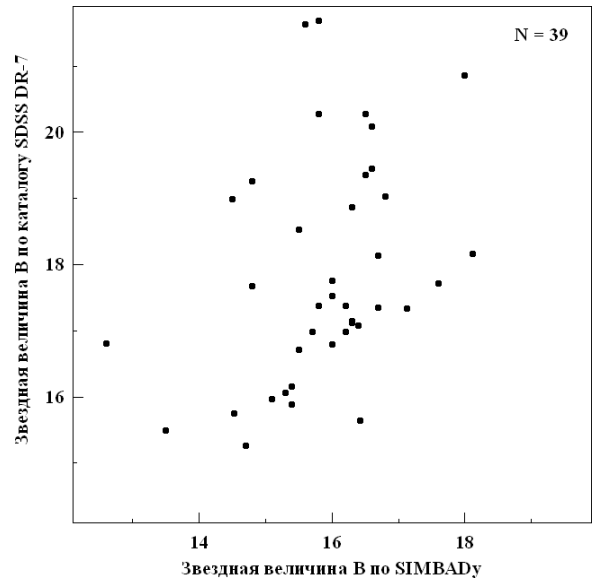

Рис. 5.

## 4 Сравнение данных SIMBAD и VSX

Для дополнительного выяснения обнаруженного несоответствия, как и в предыдущих случаях, была произведена выборка данных, вошедших в GTSh10, из Индекс каталога переменных звёзд – VSX (Ватсон, 2006). 253 объекта, имеющих *B* величины в GTSh10 были обнаружены в VSX.

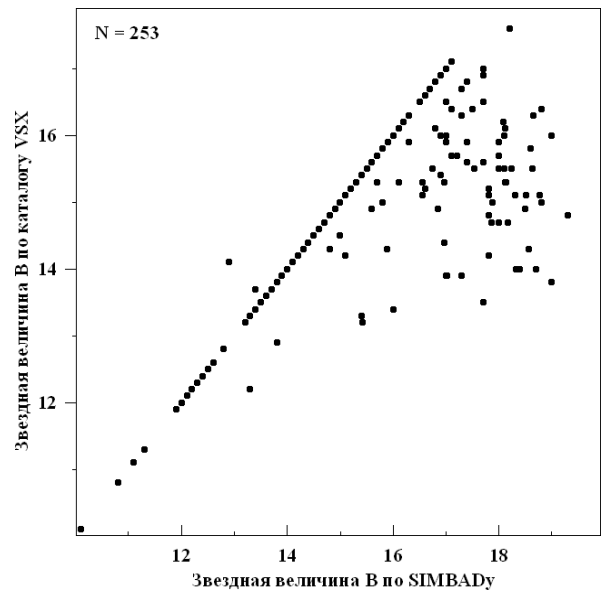

Рис. 6.

Рис. 6 иллюстрирует зависимость между блеском вспыхивающих звёзд по данным SIMBAD и в максимуме блеска по данным VSX. 52 объекта из 253 демонстрируют линейную связь в звёздных величинах. Таким образом, одним из объяснений обнаруженного при составлении GTSh10 большего блеска в сравнении со средними каталожными значениями для части объектов может служить внесение в базу данных SIMBAD звёздных величин объектов в активной фазе, т.е. во время вспышки блеска.

Очевидно, что возникшая ситуация связана со спецификой формирования базы данных SIMBAD, в которой наполнение информацией проводится путем перекрестного отождествление объектов из каталогов, списков и журнальных статей (Вингер, 2000), т.е. изначально неоднородного материала.

Анализ представления в SIMBAD данных для звёзд каталога GTSh10

## 5 Сравнение звёзд GTSh10 в минимуме по данным NOMAD и VSX

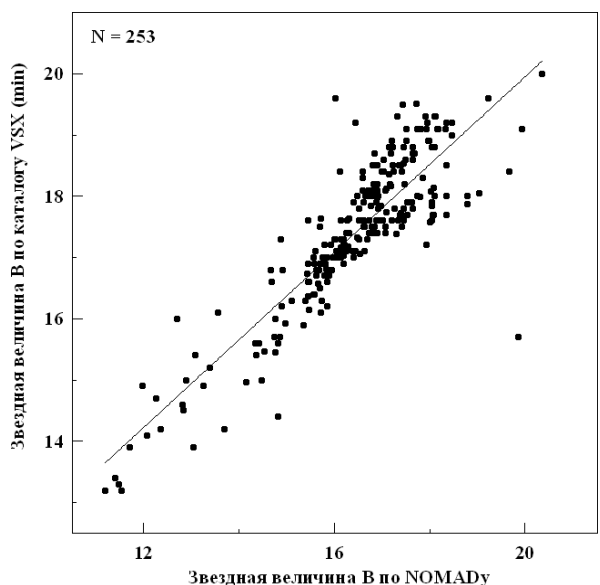

Рис. 7.

С целью повышения достоверности выводов о сходимости данных для вспыхивающих звёзд в неактивном состоянии по VSX с данными других каталогов, была исследована зависимость между минимумами блеска объектов по Индекс каталогу переменных звёзд и блеском звёзд по NOMAD.

Из рис. 7 видно, что зависимость близка к линейной (коэффициент корреляции 0.87). Разброс значений на рис. 7, также как и на рис. 3 обусловлен разными эпохами наблюдений объектов, имеющих спорадические и периодические изменения блеска. Поэтому, учитывая ориентацию созданного каталога GTSh10 на планирование наблюдений включенных в него объектов, данные о звёздных величинах были унифицированы и приведены к минимуму блеска.

## 6 Заключение

В результате проведённого исследования было определено, что для некоторых красных карликов с активностью солнечного типа, вошедших в каталог GTSh10, в базе данных SIMBAD указаны звёздные величины соответствующие максимальным значениям блеска. Это может вносить неопределённость в отождествление звёзд при проведении обзорных наблюдений. Отметим также, что некоторые из вспыхивающих звёзд (особенно в области Ориона), представленных в SIMBAD отсутствуют на самых глубоких изображениях из обзоров всего неба. Это тоже является следствием внесения в базу данных информации об объектах в стадии высокой активности. Все обнаруженные несоответствия будут учтены при подготовке дополнений к GTSh10, которые составят основу нового каталога.

При подготовке статьи активно использовались поддерживаемые Центром астрономических данных в Страсбурге приложения SIMBAD и VizieR. Автор признателен всем, кто обеспечивает их работу.

## Литература